\documentclass[12pt]{article}

\usepackage[totalheight = 23cm, totalwidth = 17cm]{geometry}
\usepackage{amssymb,amsmath,amsfonts,amsbsy,graphicx}
\usepackage[dvipsnames]{xcolor}
\usepackage[%
             colorlinks=true,urlcolor=MidnightBlue,linkcolor=MidnightBlue,citecolor=OliveGreen,
             pdfpagelabels=true,hypertexnames=true,
            plainpages=false,naturalnames=false,
             ]{hyperref}
\usepackage{epstopdf}

\def\mathbi#1{\textbf{\em #1}}

\newcommand{\mpl}{m_{\rm Pl}}

\newcommand{\calF}{{\cal F}}
\newcommand{\calL}{{\cal L}}
\newcommand{\calO}{{\cal O}}
\newcommand{\calP}{{\cal P}}
\newcommand{\calR}{{\cal R}}

\begin{document}

\begin{titlepage}

\rightline{\footnotesize{APCTP-Pre2014-011}}
\rightline{\footnotesize{CTPU-14-07}}

\begin{center}

\vskip 1cm

\Huge{Higher derivatives and power spectrum \\ in effective single field inflation}

\vskip 1cm

\large{
Jinn-Ouk Gong$^{a,b}$,
\hspace{0.1cm}
Min-Seok Seo$^c$
\hspace{0.1cm} and \hspace{0.1cm}
Spyros Sypsas$^a$
}

\vskip 1.2cm

\small{\it
$^{a}$Asia Pacific Center for Theoretical Physics, Pohang 790-784, Korea
\\
$^{b}$Department of Physics, Postech, Pohang 790-784, Korea
\\
$^{c}$Center for Theoretical Physics of the Universe, Institute for Basic Science, Daejeon 305-811, Korea
}

\vskip 1.2cm

\end{center}

\begin{abstract}

We study next-to-leading corrections to the effective action of the curvature perturbation obtained by integrating out the coupled heavy isocurvature perturbation. These corrections result from including higher order derivative operators, weighted by the mass scale of the heavy physics, in the effective theory expansion. We find that the correction terms are suppressed by the ratio of the Hubble parameter to the heavy mass scale. The corresponding corrections to the power spectrum of the curvature perturbation are presented for a simple illustrative example.

\end{abstract}

\end{titlepage}

\setcounter{page}{0}
\newpage
\setcounter{page}{1}

\section{Introduction}
\label{sec:intro}

Primordial cosmic inflation~\cite{inflation} is the leading paradigm that explains the observed homogeneity and isotropy of the universe as seen in the cosmic microwave background (CMB). At the same time, during inflation small quantum fluctuations are stretched to cosmic scales and become the seed of the subsequent large scale inhomogeneities such as the temperature fluctuations in the CMB~\cite{Mukhanov:2005sc}. While most recent observations are consistent with the predictions of inflation~\cite{planck}, the microphysical details of the actual inflationary model are however still speculative. This is because of our ignorance of the ultraviolet (UV) complete theory in which the inflaton and other relevant degrees of freedom are embedded. In this respect, the bottom-up effective field theory (EFT) approach to inflation~\cite{Cheung:2007st,Weinberg:2008hq} is very attractive since it allows for a systematic control over our ignorance, and hence enables us to address the generic impact of elusive UV physics on the low energy observables~\cite{Burgess:2014lza}.

EFT consists of the derivative expansion with respect to a cutoff scale $M$ that represents the unknown physics below which a canonical, effectively single field description of inflation is legitimate~\cite{noheavy}. This scale $M$ can be considered as the mass scale heavier than the characteristic scale during inflation, viz. the Hubble parameter $H$, so that UV physics decouples at the leading EFT. The effects of heavy physics, however, permeate the couplings of the derivative expansion and thus manifest themselves in the low energy EFT that describes otherwise canonical single field inflation. The leading couplings are parametrized by the effective speed of sound $c_s$ that can exhibit departure from 1, stemming from transient strong coupling of certain operators of the EFT~\cite{noheavy,soundspeed}. The associated observational effects are the non-trivial, scale dependent oscillations in the power spectrum of the curvature perturbation and other correlated higher order correlation functions~\cite{corr-corr}.

Thus, we are naturally led to consider the next-to-leading effects of the derivative couplings of the EFT. The reason is twofold. First, it is a natural extension beyond the leading derivative expansion. This includes subtle theoretical considerations and reminds us of what may be easily overlooked in applications of field theory. Moreover, as we approach the cutoff scale $M$, the heavy modes are invoked and the sub-leading corrections become more and more important. Thus incorporating further corrections is indispensable to more accurately estimate the resulting observable signatures.

One technical subtlety is that one is forced to deal with higher order spacetime derivatives. As noted by Ostrogradsky~\cite{Ostrogradsky}, theories with arbitrary derivatives are plagued by instabilities since in such cases a variable and its time derivative do not satisfy the canonical commutation relation any longer, but they are formally treated as commuting independent variables. In order to cure such potential instabilities in the EFT context, one has to recall that the theory is represented as a series over a dimensionless small parameter, usually the ratio of the characteristic scale over the heavy one, and the solutions of the theory should also respect such an expansion. Therefore, one may treat higher time derivatives that appear in sub-leading interaction terms by making use of the leading equation of motion. This procedure is valid at the next-to-leading order, since at this level it is equivalent to a field redefinition including a field and its derivatives in the Ostrogradsky formalism.

This article is outlined as follows: in Section~\ref{sec:general}, we give a general discussion on how to deal with higher derivative terms in field theory. In Section~\ref{sec:power}, we consider the quadratic effective single field action of the curvature perturbation when a heavy degree of freedom is integrated out. By removing higher order derivatives applying the formalism discussed in Section~\ref{sec:general}, we find that the effective speed of sound is modified, leading to a dispersion relation with quartic momentum dependence. We then compute the change in the power spectrum of the curvature perturbation, and find that the next-to-leading corrections are appreciable when the mass scale of the heavy modes is not too large compared to the Hubble scale. We then conclude in Section~\ref{sec:conc}.

\section{General arguments}
\label{sec:general}

In this section, we consider how to treat higher derivatives in the low energy EFT, consisting of the derivative expansion with respect to a cutoff scale $M$. As an example, one may take the four-Fermi interaction where momentum transfer between two charged currents is suppressed by the $W$ boson mass~\cite{Georgi:1985kw}. Another example is the chiral perturbation theory~\cite{Chiralpert}, where mesons consisting of light quarks, $u$, $d$ and $s$, are interpreted as massless Goldstone bosons resulting from spontaneous breaking of global $SU(3)_L \times SU(3)_R$ upon neglecting the light quark masses. In this case, the theory is described by derivatives of the $SU(3)$-valued meson field suppressed by the pion decay constant. In general, studying an EFT to a specific accuracy invites us to consider a theory with spacetime derivatives of a certain higher order.

The manipulation of higher derivatives is known as the Ostrogradsky formalism~\cite{Ostrogradsky}. If a Lagrangian depends on a variable $q$ and its derivatives up to order $N$,
\begin{equation}\label{thL}
L\left( q^{(0)}\equiv q, q^{(1)}, \cdots, q^{(N)} \right) \, ,
\end{equation}
where $q^{(n)}\equiv(d/dt)^n q$, we have $N$ independent variables and their conjugate momenta,
\begin{align}
Q^n & = q^{(n-1)} \, ,
\\
P_n & = \sum_{i=n}^{N}\Big(-\frac{d}{dt}\Big)^{i-n}\frac{\partial L}{\partial q^{(i)}} \, ,
\end{align}
where $n$ runs from 1 to $N$. With these canonical variables, we can construct the Hamiltonian as $H=\sum_{n=1}^N P_n {\dot Q}^n-L$, from which the Hamiltonian equations of motion may be derived in the usual way. The point here is that in the Ostrogradsky formalism, all but the highest derivative of the variable are treated as independent degrees of freedom. Especially, $q$ and ${\dot q}$ do not form a canonical conjugate pair any longer. This gives rise to a dangerous feature of theories with higher derivatives, known as the Ostrogradsky instability. Since quadratic time derivatives are not guaranteed, arbitrary negative kinetic energy makes the energy unbounded from below in general.

Of course, EFT belongs to the special case of stable energy since the UV complete theory in which it is embedded is assumed to be a healthy one~\cite{stableE}. Moreover, in such a framework we want to  keep a field and its time derivative as the canonical pair and treat higher derivative effects as perturbations.  For this purpose, it was suggested to replace the leading higher derivative terms with non-derivative ones using the equations of motion obtained from leading interactions~\cite{onshellEFT}. Even though this is useful for next-to-leading order (NLO) calculations, this method is somehow speculative, since it looks as if classical dynamics, represented by equations of motion, yield some features of the whole dynamics deduced from the Lagrangian, including quantum effects. In fact, it is known that the manipulation of higher derivatives through the leading equation of motion does not hold in the next-to-next-to-leading order (NNLO) and beyond~\cite{Scherer:1994wi}. This is because the replacement of higher derivatives at NLO using the equation of motion is equivalent to a field redefinition involving time derivatives of the field as well as the field itself, and another field redefinition aiming at the removal of higher derivatives at NNLO cannot anymore be thought of as stemming from the use of the equation of motion~\cite{GrosseKnetter:1993td}. Note that we are free to include time derivatives in the redefinition precisely because in the Ostrogradsky formalism fields and and their (higher) derivatives are just independent, commuting variables. Hence, a local transformation that involves these variables, or ``coordinates'', preserves the physical content of the system under consideration, even if the transformation contains derivatives.

For clarity, let us consider a simple toy model~\cite{Weinberg:2008hq}
\begin{equation}\label{toyL}
\calL = -\frac{1}{2} \left[ \partial_\mu\phi\partial^\mu\phi + m^2\phi^2 + \frac{\left(\Box\phi\right)^2}{M^2} \right] + J\phi \, ,
\end{equation}
where $M \gg m$ is a very large mass scale, and $J$ is an external current. Treating the higher order derivative term $\left(\Box\phi\right)^2/M^2$ as a perturbation, we obtain the vacuum persistence amplitude as
\begin{equation}\label{Gamma}
\Gamma = i \int d^4k \frac{|J(k)|^2}{k^2+m^2} \left[ 1 - \frac{k^4}{M^2 \left( k^2+m^2 \right)} \right] \, .
\end{equation}
Alternatively, we can make use of the equation of motion derived from the leading terms,
\begin{equation}
\Box\phi = m^2\phi - J \, ,
\end{equation}
to eliminate $\Box\phi$ at $\calO(M^{-2})$ and obtain an action without higher derivatives, namely
\begin{equation}\label{derfreeL}
\calL = -\frac{1}{2} \left( \partial_\mu\phi\partial^\mu\phi + m^2\phi^2 + \frac{m^4}{M^2}\phi^2 \right) + \left( 1 + \frac{m^2}{M^2} \right) J\phi - \frac{J^2}{2M^2} \, ,
\end{equation}
which is equivalent to the original theory to $\calO(M^{-2})$. This can be shown by checking that the two Lagrangians \eqref{toyL} and \eqref{derfreeL} give rise to the same $\Gamma$ given by \eqref{Gamma}. As mentioned above, this amounts to a field redefinition including derivatives, i.e.
\begin{equation}\label{toyredef}
\phi \to \phi + \frac{1}{2M^2} \left( \Box\phi + m^2\phi - J \right) \, .
\end{equation}

In order to make the discussion complete, we now comment on the elimination of higher derivatives beyond the leading order. In this case, naively using the equation of motion is insufficient, since the resulting vacuum persistence amplitude differs from the one obtained from the original theory. To be more concrete, let us return to the toy Lagrangian \eqref{toyL}. Before we proceed,  we should note that \eqref{toyL} is truncated at $\calO(1/M^2)$. Therefore, we should extend it by including possible $\calO(1/M^4)$ terms as
\begin{equation} \label{toy-L-ext}
\calL = -\frac{1}{2} \left[ \partial_\mu\phi\partial^\mu\phi + m^2\phi^2 + \frac{\left(\Box\phi\right)^2}{M^2} \right] + J\phi + \frac{1}{8M^4} \left[ c_1\Box\phi\Box^2\phi + c_2m^2\left(\Box\phi\right)^2 + c_3m^4\phi\Box\phi \right] \, .
\end{equation}
Then, in addition to the field redefinition \eqref{toyredef}, we have to consider the following $\calO(1/M^4)$ terms:
\begin{align} \label{extended-redef}
\frac{1}{8M^4} & \left[ (-c_1+3)\Box^2\phi + (-c_1-c_2+6)m^2\Box\phi + (-c_1-c_2-c_3+7)m^4\phi \right.
\nonumber\\
&\left.+ (2c_1+c_2-11)m^2J + (c_1-5)\Box J \right] \, . 
\end{align}
This expression leads to an effective Lagrangian without higher derivatives on $\phi$ as\footnote{Note that eliminating the $c_3[m^4/(8M^4)]\phi\Box\phi$ term in \eqref{toy-L-ext} can be regarded as a wavefunction renormalization.}
\begin{align}\label{toyL2}
\calL = & -\frac{1}{2} \left[ \partial^\mu\phi\partial_\mu\phi + m^2\phi^2 + \frac{m^4}{M^2}\phi^2 - (c_1+c_2+c_3-8)\frac{m^6}{4M^4}\phi^2 \right]
\nonumber\\
& + \left[ 1 + \frac{m^2}{M^2} + (20-3c_1-2c_2-c_3)\frac{m^4}{8M^4} \right]J\phi 
\nonumber\\
& - \frac{1}{2M^2}
{J\left\{ 1 - \left[ (c_1-4)\Box + (2c_1+c_2-12)m^2 \right] \frac{1}{4M^2} \right\}J} \, ,
\end{align}
which results in the same vacuum persistence amplitude as the one obtained from the original Lagrangian,
\begin{align}
\Gamma = & \int \frac{d^4k}{(2\pi)^4} \frac{|J(k)|^2}{k^2+m^2} \left[ 1 - \frac{k^4}{M^2\left( k^2+m^2 \right)} \right.
\nonumber\\
& \left. \hspace{3.2cm} + \frac{(-c_1+4)k^8 + (-c_1+c_2)m^2k^6 + (c_2-c_3)m^4k^4 - c_3m^6k^2}{4M^4\left( k^2+m^2 \right)^2} \right] \, .
\end{align}

Note that in obtaining \eqref{toyL2}, we have concentrated only on $\Box^n\phi$. If we were to remove $\Box J$ as well, we would have to add $\delta\phi=[c_J/(8M^4)]\Box J$ to \eqref{extended-redef} and impose $c_J=-1$, in which case $c_1$, $c_2$ and $c_3$ would also be determined. However, this is not what we expect from EFT because these coefficients should be determined from the physics beyond the scale $M$, which in general does not force the condition $c_J=-1$. In fact, we do not need to eliminate $\Box J$ terms because $J$ here is just a background quantity not subject to dynamics. Moreover, the surviving term $[(c_1-4)/(8M^4)]J\Box J$ does not destabilize the energy since the Lagrangian is only valid for $k^2\ll M^2$.

\section{Power spectrum with higher derivatives}
\label{sec:power}

Having discussed general issues of higher derivative terms in field theory, we now turn to the subject of our interest, the corrections to the power spectrum of the curvature perturbation due to higher derivative terms induced from integrating out a heavy isocurvature mode. We begin with a simple two-field action
\begin{equation}
S = \int d^4x \sqrt{-g} \left[ \frac{\mpl^2}{2}R - \frac{1}{2}g^{\mu\nu}\partial_\mu\phi^a\partial_\nu\phi_a - V(\phi) \right] \, ,
\end{equation}
and choose the so-called comoving gauge, where $\phi^a(t,\mathbi{x})$ and the spatial metric $h_{ij}(t,\mathbi{x})$ are respectively written as~\cite{noheavy}
\begin{align}
\phi^a(t,\mathbi{x}) & = \phi_0^a(t) + N^a(t)\calF(t,\mathbi{x}) \, ,\label{Eq:phi}
\\
h_{ij}(t,\mathbi{x}) & = a^2(t)e^{2\calR(t,\mathbi{x})}\delta_{ij} \, \label{Eq:metric}.
\end{align}
Here, $N^a$ is the unit vector normal to the background trajectory, and $\calF$ denotes the deviation from $\phi^a_0(t)$, i.e. the heavy isocurvature perturbation, while $\calR$ denotes the adiabatic perturbation of the hypersurfaces under this gauge condition, viz. the comoving curvature perturbation.

In order to investigate the dynamics, we proceed in the standard way of performing an Arnowitt-Deser-Misner decomposition of the metric~\cite{Arnowitt:1962hi}. Plugging (\ref{Eq:phi}) and (\ref{Eq:metric}) into the action, imposing the lapse and shift constraints and expanding, the action at quadratic order reads
\begin{align}
\label{totalS}
S = & \int d^4x a^3\epsilon\mpl^2 \left[ \dot\calR^2 - \frac{(\nabla\calR)^2}{a^2} \right] 
+ \int d^4x \left\{ \frac{a^3}{2} \left[ \dot\calF^2 - \frac{(\nabla\calF)^2}{a^2} - M^2F^2 \right] - 2a^3 \frac{\dot\theta\dot\phi_0}{H} \calF\dot\calR \right\}
\nonumber\\
\equiv & S_\calR + S_\calF + S_\text{int} \, ,
\end{align}
where $\dot\phi_0 \equiv \sqrt{\dot\phi^a\dot\phi_a}$ is the rapidity of the scalar field's vacuum expectation value, $\epsilon \equiv -\dot{H}/H^2 = (\dot\phi_0/H)^2/(2\mpl^2)$ is the slow-roll parameter, $\dot\theta \equiv N^aV_a/\dot\phi_0$ is the angular velocity for the trajectory and $M^2 \equiv N^aN^bV_{ab} - \dot\theta^2$. With such an interaction term, the formal solution of $\calF$ is written as
\begin{equation}\label{eq:Fsol}
\calF = \left( \Box-M^2 \right)^{-1} \frac{2\dot\theta\dot\phi_0}{H}\dot\calR \, ,
\end{equation}
where 
\begin{equation}
\Box \equiv \left. \frac{1}{\sqrt{-g}} \partial_\mu \left( \sqrt{-g}g^{\mu\nu}\partial_\nu \right) \right|_\text{background} = -\frac{d^2}{dt^2} - 3H\frac{d}{dt} + \frac{\Delta}{a^2} \, .
\end{equation}
Here, $\left( \Box - M^2 \right)^{-1}$ should be understood as the inverse operator of $\Box - M^2$. Expanding this non-local derivative operator as a power series in $\Box/M^2$ we may represent the resulting action as a sum of local operators, namely
\begin{equation} \label{Seff-int}
S_\calF + S_\text{int} = \int d^4x \left[ a^3\epsilon\mpl^2 \frac{4\dot\theta^2}{M^2}\dot\calR^2 + \frac{2a^3}{M^4} \frac{\dot\theta\dot\phi_0}{H}\dot\calR \Box \left( \frac{\dot\theta\dot\phi_0}{H}\dot\calR \right) + \calO\left(M^{-6}\right) \right] \, .
\end{equation}

The first term induces the leading effect in $\Box/M^2$, which can be identified with a non-trivial speed of sound~\cite{noheavy,soundspeed}
\begin{equation}\label{leading_cs}
\frac{1}{c_s^2} = 1 + \frac{4\dot\theta^2}{M^2} \, .
\end{equation}
After a series of partial integrations the effective action including next-to-leading corrections, given by the second term in \eqref{Seff-int}, reads 
\begin{equation} \label{Seff-int-1}
S_\calF + S_\text{int} = \int d^4x a^3\epsilon\mpl^2 \left( \frac{1}{c_s^2}-1 \right) \dot\calR^2 + \int d^4x a^3\epsilon\mpl^2 \left[ \tilde{c}_0^2\dot\calR^2 + 4 \frac{\dot\theta^2}{M^4} \left( \ddot\calR^2 - \dot\calR \frac{\Delta}{a^2}\dot\calR \right) \right] \, ,
\end{equation}
where
\begin{align}
\tilde{c}_0^2 & = \frac{H^2}{M^2} \left( \frac{1}{c_s^2}-1 \right) \left[ -\frac{3}{2}\eta + (-3+\epsilon-\eta-t+2m)(t-2m) + \frac{\epsilon\eta}{2} - \frac{\eta^2}{4} - 4m^2 - \frac{\dot\eta}{2H} - \frac{\dot{t}}{H} + 2\frac{\dot{m}}{H} \right] \, ,
\end{align}
with $\eta \equiv \dot\epsilon/(H\epsilon)$, $t \equiv {\ddot\theta}/(H{\dot \theta})$ and $m \equiv {\dot M}/(HM)$. Note that $m$ and especially $t$ need not be small in this context since they obstruct neither slow-roll nor the validity of the EFT~\cite{validity,Cespedes:2012hu}. Nevertheless, the adiabaticity condition ${\ddot \theta}/\left(M{\dot \theta}\right) \ll 1$~\cite{Cespedes:2012hu} implies that $t \lesssim M/H$ should be respected. As corrections under consideration are important when $H/M$ is not negligible, $t$ cannot be arbitrarily large.
Another widely adopted parameter is $s \equiv \dot{c}_s/(Hc_s)$, which is related to $t$ and $m$ via
\begin{equation}
s = \left( c_s^2-1 \right)(t-m) \, .
\end{equation}

Now, as discussed in the previous section, in order to remove the $\ddot{\cal R}^2$ term from \eqref{Seff-int-1} we should impose the equation of motion derived from the {\em leading} terms in the $\Box/M^2$ expansion, i.e.
\begin{equation}\label{Requation}
\frac{c_s^2}{a^3\epsilon}\frac{d}{dt} \left( \frac{a^3\epsilon}{c_s^2}\dot\calR \right) - c_s^2\frac{\Delta}{a^2}\calR = 0 \, ,
\end{equation}
with $1/c_s^2$ being given by \eqref{leading_cs}. From \eqref{Requation}, $\ddot\calR$ can be replaced with
\begin{equation}\label{ddotR}
-\ddot\calR = \frac{c_s^2}{a^3\epsilon} \frac{d}{dt} \left( \frac{a^3\epsilon}{c_s^2} \right) \dot\calR - c_s^2\frac{\Delta}{a^2}\calR = H(3+\eta-2s)\dot\calR - c_s^2\frac{\Delta}{a^2}\calR \, .
\end{equation}
Thus, the higher order derivative term $\ddot\calR^2$ in addition to correcting the speed of sound via the $\dot\calR^2$ and $(\nabla\calR)^2$ contributions, also adds a quartic momentum dependence to the dispersion relation through the $(\Delta\calR)^2$ term. Substituting \eqref{ddotR} in \eqref{Seff-int-1} and after several partial integrations, we obtain
\begin{equation}
S_\calF + S_\text{int} = \int d^4x a^3\epsilon\mpl^2 \left( \frac{1}{c_s^2}-1 \right) \dot\calR^2 + \int d^4x a^3\mpl^2\epsilon \left[ c_0^2\dot\calR^2 - c_2^2\frac{(\nabla\calR)^2}{a^2} - \frac{c_4^2}{H^2} \frac{(\Delta\calR)^2}{a^4} \right] \, ,
\end{equation}
where the new coefficients beyond $c_s^2$ induced by higher derivatives are
\begin{align}
c_0^2 & = \tilde{c}_0^2 + \frac{H^2}{M^2} \left( \frac{1}{c_s^2}-1 \right) (3+\eta-2s)^2 \, ,
\\
c_2^2 & = \frac{H^2}{M^2} \left( 1-c_s^2 \right) \left\{ (1-\epsilon+\eta+2t-4m) \left[ 3c_s^2 - 1 + c_s^2\eta - 2\left( c_s^2-\frac{1}{2} \right) s + t - 2m \right] \right.
\nonumber\\
& \left. \hspace{3.3cm} + 2c_s^2s(3+\eta-2s) + c_s^2\frac{\dot\eta}{H} + \frac{\dot{t}}{H} - 2\frac{\dot{m}}{H} + \left( 1-2c_s^2 \right)\frac{\dot{s}}{H} \right\} \, ,
\\
c_4^2 & = \frac{H^2}{M^2} \left( 1-c_s^2 \right)^2 \, .
\end{align}
Note that these dimensionless coefficients have the common suppression factor $H^2/M^2$ and vanish when $c_s^2 \to 1$, or equivalently in the limit $M\to\infty$ or $\dot\theta \to 0$, in agreement with our intuition. Finally, adding $S_\calR$ given in \eqref{totalS}, the final quadratic effective action for $\calR$ including next-to-leading expansion in $\Box/M^2$ reads
\begin{equation}\label{Sfin}
S = \int d^4x a^3\mpl^2\epsilon \left[ \left( \frac{1}{c_s^2} + c_0^2 \right)\dot\calR^2 - \left( 1 + c_2^2 \right) \frac{(\nabla\calR)^2}{a^2} - \frac{c_4^2}{H^2} \frac{(\Delta\calR)^2}{a^4} \right] \, .
\end{equation}

The dispersion relation can be read off from the action \eqref{Sfin} as 
\begin{equation}\label{eq:dispersion}
\omega^2 = \frac{\left(1+c_2^2\right)c_s^2}{1+c_0^2c_s^2} p^2 + \frac{c_4^2c_s^2}{1+c_0^2c_s^2} \frac{p^4}{H^2} \, ,
\end{equation}
where $p \equiv k/a$ denotes the physical momentum. The first term of \eqref{eq:dispersion} gives the effective speed of sound including, in addition to the leading result \eqref{leading_cs}, slow-roll and $H^2/M^2$ suppressed corrections. In order to see what the second term of \eqref{eq:dispersion} signifies, it is instructive to recall that the EFT under consideration is only valid when we can expand the non-local operator in \eqref{eq:Fsol} in terms of $\Box/M^2$, viz. $\omega^2 < p^2 + M^2$. The energy scale where this relation breaks down and the UV degree of freedom becomes dynamical is given by $\Lambda_{\rm UV} \sim M/c_s$~\cite{Gwyn:2012mw}. The second term of \eqref{eq:dispersion} can be written as $p^4/\Lambda_{\rm UV}^{'2}$, where $\Lambda_{\rm UV}^{'2}=\Lambda_{\rm UV}^{2} \left[ 1+\mathcal{O}(\epsilon,H^2/M^2) \right]$ denotes the UV scale with slow-roll and $H^2/M^2$ corrections from higher order derivatives.

A similar dispersion relation, with a quadratic-quartic structure, was also found in~\cite{Gwyn:2012mw,Baumann:2011su} (see also \cite{new-phys}) and it can be easily seen that \eqref{eq:dispersion} reduces to the one studied in these works in the limit where slow-roll and $H^2/M^2$ corrections are dropped. However, in~\cite{Gwyn:2012mw} an alternative EFT expansion was used, based on the assumption of a sufficiently small speed of sound, while in \cite{Baumann:2011su} slow-roll and heavy field corrections were neglected. Due to these differences, the equation of motion for the interaction picture field reported there is rather different from the one derived from \eqref{Sfin}, leading to a different scaling of the solution\footnote{The solution is still a Hankel function multiplied with a momentum prefactor such that it is scale invariant in the super-horizon limit. However, because of the different expansion and the slow-roll and heavy field corrections, the order of the Hankel function and consequently the momentum factor are different from \cite{Gwyn:2012mw,Baumann:2011su}.}. In particular, one does not have to confine oneself in the so-called new physics regime of~\cite{Gwyn:2012mw,Baumann:2011su,new-phys}, since the effective theory can now be solved throughout its full validity window. We expect that the different scaling of the quadratic operators will also result in distinct momentum dependence of the cubic operators and consequently in distinct integrands in the computation of three-point correlators. It would be interesting to compare the non-Gaussian signatures of the two EFT expansions, i.e. \eqref{Sfin} and the one studied in ~\cite{Gwyn:2012mw,Gwyn:2014doa} to higher order in slow-roll corrections, along the lines of~\cite{Burrage:2011hd}.

Having found the quadratic action \eqref{Sfin}, we may now proceed to compute the corresponding change in the power spectrum $\calP_\calR$. With $\calP_\calR = H^2/(8\pi^2\epsilon\mpl^2)$ being the featureless, flat power spectrum derived from the free part of the action, we may treat $S_\calF + S_\text{int}$ as a perturbation and use the standard de Sitter mode functions and the in-in formalism~\cite{in-in}. Then we find the resulting change in the power spectrum as
\begin{align}\label{eq:DeltaP}
\frac{\Delta\calP_\calR}{\calP_\calR} & = \kappa \int_0^\infty d\tau \left( \frac{1}{c_s^2}-1 \right) \sin(2\kappa\tau) 
\nonumber\\
& \hspace{0.5cm} + \kappa \int_0^\infty d\tau c_0^2 \sin(2\kappa\tau) + \frac{1}{\kappa} \int_0^\infty d\tau c_2^2 \left[ -\frac{\sin(2\kappa\tau)}{\tau^2} + \frac{2\kappa}{\tau}\cos(2\kappa\tau) + \kappa^2\sin(2\kappa\tau) \right] 
\nonumber\\
& \hspace{0.5cm} - \kappa \int_0^\infty d\tau c_4^2 \left[ -\sin(2\kappa\tau) + 2\kappa\tau\cos(2\kappa\tau) + \kappa^2\tau^2\sin(2\kappa\tau) \right] \, ,
\end{align}
where $\kappa \equiv k/k_\star$ and $\tau \equiv \tilde\tau/\tilde\tau_\star$, with $\tilde\tau = \int dt/a$ being the conformal time and $\star$ denoting a convenient reference. The first term of \eqref{eq:DeltaP} represents the leading correction generated by changes in the speed of sound, while the following three terms arise from higher derivatives. The corresponding changes in the spectral index $n_\calR = 1 - 2\epsilon - \eta$ are easily computed to give, to leading order in slow-roll parameters,
\begin{equation}
\Delta n_\calR = - s - 3s\frac{H^2}{M^2} \, .
\end{equation}
Note that the same result can be reached by solving the equation of motion for $\calR$ derived from \eqref{Sfin}.

\begin{figure}[t]
 \begin{center}
  \includegraphics[width=15cm]{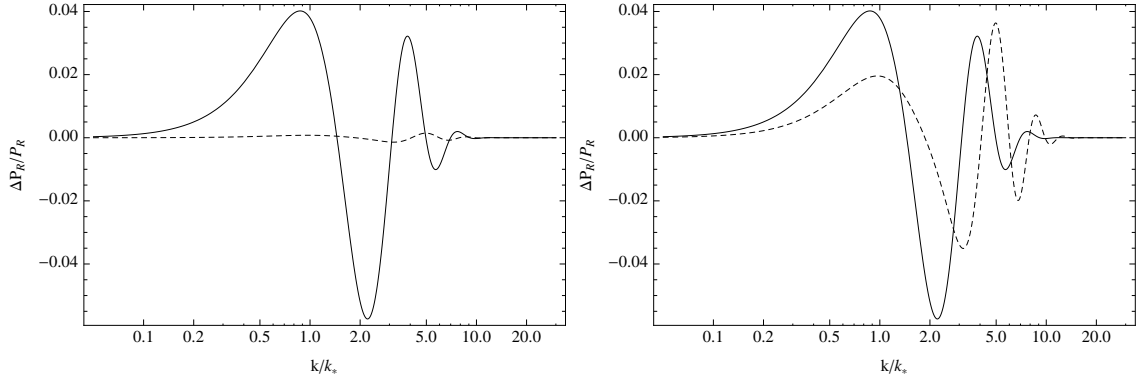}
  \vspace{-1em}
 \end{center}
 \caption{Plot of $\Delta\calP_\calR/\calP_\calR$, with the leading (solid) and next-to-leading corrections (dotted) shown separately. In the left and right panels we set $H^2/M^2 = 1/300$ and $1/12$ respectively, while fixing $c_\text{max}=1/12$.}
 \label{fig:example}
\end{figure}

In Figure~\ref{fig:example}, we plot the change in the power spectrum \eqref{eq:DeltaP} for an illustrative example where the speed of sound is an analytic function of the number of $e$-folds $N$ as \cite{noheavy}
\begin{equation}
\frac{1}{c_s^2} = 1 + c_\text{max} \cosh^{-4} \left[ \frac{2(N-N_\star)}{\Delta{N}} \right] \, ,
\end{equation}
where $c_\text{max}$ is the maximum departure of $c_s^2$ from 1 peaked at $N_\star$, and $\Delta{N}$ is the number of $e$-folds during which $c_s^2$ deviates from 1. For simplicity and to isolate the genuine effects from a turning trajectory, we assume that $H$ and $M$ are constant. While current observations constrain $c_\text{max} \lesssim 10^{-1}$~\cite{planck}, we can vary the ratio of the cutoff scale $M$ to $H$. As we can see, if $H^2/M^2 \ll 1$ the leading correction is sufficient to describe the oscillatory features in the power spectrum. But as $H^2/M^2$ increases, the next-to-leading corrections coming from higher derivative terms become more and more important.

\section{Conclusions}
\label{sec:conc}

In this article, we have studied the next-to-leading corrections to the quadratic action of the curvature perturbation $\calR$, obtained by integrating out a heavy isocurvature perturbation $\calF$ characterized by a mass scale $M$. These corrections are coming from the expansion in $\Box/M^2$ of the effective theory. This requires special care for the higher derivative terms, which can be replaced by a systematic field redefinition that is equivalent, at the NLO expansion, to using the equation of motion. The resulting effective action includes operators suppressed by $H^2/M^2$, which induce corrections to the speed of sound as well as a quartic contribution to the dispersion relation. The corresponding change in the power spectrum $\calP_\calR$ and the spectral index $n_\calR$ is appreciable as we approach the heavy mass scale $M$.

\subsection*{Acknowledgements}

We thank Subodh Patil and Masahide Yamaguchi for helpful conversations.
JG and SS acknowledge the Max-Planck-Gesellschaft, the Korea Ministry of Education, Science and Technology, Gyeongsangbuk-Do and Pohang City for the support of the Independent Junior Research Group at the Asia Pacific Center for Theoretical Physics. JG and SS are also supported by a Starting Grant through the Basic Science Research Program of the National Research Foundation of Korea (2013R1A1A1006701).

\end{document}